\documentclass{aa}
\usepackage{graphics}

\begin{document}

\thesaurus{(05.05.2, 08.02.4, 08.09.2, 13.25.5)}

\title{An Improved Orbital Ephemeris for Cygnus X-1}

\author{C. Brocksopp\inst{1}
  \and A.E.Tarasov\inst{2} 
  \and V.M.Lyuty\inst{3} 
  \and P.Roche\inst{1}}

\institute{Astronomy Centre, University of Sussex, Falmer, Brighton, U.K.
  \and Crimean Astrophysical Observatory, Ukraine, 334413, Nauchny, Crimea, Ukraine 
  \and Sternberg Astronomical Institute, Universitetskii pr.13, Moscow, 119899 Russia}

\mail{cb@star.cpes.susx.ac.uk}

\date{received date / Accepted date}

\maketitle

\begin{abstract}

By accumulating all available radial velocity data and combining it with our own spectroscopy we pres-\\ent a new and improved ephemeris for Cygnus X-1 (V1357 Cygni/HDE226868). We initially calculate an orbital period of $5.599847\pm0.000018$ days and then, by including the more recent radial velocities of LaSala et al. (1998) and Sowers et al.(1998) we refine this to $5.599829\pm0.000016$ days. Using  27 years' worth of photometry, we also calculate a photometric period of $5.599836\pm0.000024$ --- remarkably similar to the spectroscopic one.


\end{abstract}

\section{Introduction}

Cygnus X-1 has become perhaps the most extensively studied X-ray binary, particularly due to its providing the first observational evidence for black holes. Webster \& Murdin (1972) and Bolton (1972) independently associated the X-ray source with HDE 226868, found an orbital period of 5.6 days and calculated the first mass function to indicate a compact object more massive than 3 $M_{\odot}$, the generally accepted maximum mass of a neutron star. The values of the masses of the two components have been refined a number of times, most recently by Herrero et al. (1995) giving 17.8 $M_{\odot}$ for the supergiant companion star and 10.1 $M_{\odot}$ for the black hole.

A number of periods have been reported in the various lightcurves of Cygnus X-1. These include 294 days in the X-rays (Priedhorsky, Terrell \& Holt 1983) and optical photometry (Kemp et al. 1987), 150 days in the radio (Pooley, Fender \& Brocksopp 1998) 78/39 days in UV polarimetry (Kemp, Herman, Rudy \& Barbour 1977), optical photometry (Kemp, Herman \& Barbour 1978) and 4.5 years in optical photometry (Wilson \& Fox 1981).

However, the orbital period is the only one that is confirmed in all wavebands and by most authors. It has been detected in the X-ray data of {\it Ariel V}/ASM (Holt, Kaluzienski, Boldt \& Serlemitsos 1976), {\it RXTE}/ASM\\ (Zhang, Robinson \& Cui 1996) and {\it CGRO}/BATSE (Paciesas et al. 1997) and also in the radio (Pooley, Fender \& Brocksopp 1998), optical photometry (Volushina, Lyuty \& Tarasov 1997), optical spectroscopy (e.g. Gies \& Bolton 1982) and infrared (Leahy \& Ananth  1992; Nadzhip et al. 1996).

The spectroscopic ephemeris has been revised many times using a variety of lines. The results of Gies \& Bolton (1982) have been used ($P_{orb}=5.59974\pm0.0008$ days) as the definitive values for over a decade and there has been little disagreement; while a variable orbital period was tentatively suggested (Ninkov, Walker \& Yang 1987) this has never been confirmed and appears unlikely. It is only recently that a new ephemeris has been calculated ($P_{orb}=5.5998 \pm 0.0001$ days) and this is within the errors of the Gies \& Bolton (1982) result (LaSala et al. 1998). A more accurate result is given as $P_{orb}=5.59977 \pm 0.00002$ days (Sowers et al. 1998) which also takes Hipparcos photometrical data into account.

The spectrum is that of a typical O9 supergiant with moderate H$\alpha$ emission, presumably coming from the supergiant (e.g. Brucato \& Zappala 1974), and also He\,{\sc ii}\,$\lambda$4686 emission. Both of these emission lines are superimposed on absorption components and, by distinguishing the two components, it has been shown by a number of authors (e.g. Aab 1983a) that the He\,{\sc ii} emission component is $\sim 115-120 ^{\circ}$ out of phase with the He\,{\sc ii} absorption component and the supergiant. As a result of this it is thought that the He\,{\sc ii} emission originates in the accretion stream.

Neither of these absorption components are suitable for measuring radial velocity due to the inaccuracies involved in the removal of the superimposed emission. Instead, as previously, we turn to the He\,{\sc i} absorption lines which are uncontaminated by emission.

\section{Observations and Discussion}

\subsection{Spectroscopy}

Our spectra were obtained in June/July 1997 using the Coud\'{e} spectrograph of the Crimean Astrophysical Observatory's 2.6 metre telescope. The detector was a CDS9000 (1024$\times$256 pixels) CCD array. All observations were made in the second order of a diffraction grating with reciprocal dispersion of 3\AA/mm and resolution of 25000. The typical exposure time for each spectrum totalled 1.5 hours resulting in a S/N of $\sim$100.

With a spectral width of 60\AA, our 20 spectra centred on He\,{\sc ii}\,$\lambda4686$ also included the He\,{\sc i}\,$\lambda4713$ line. This is the line we have used for our radial velocity studies, although other sources have used alternative helium lines.

The spectra were reduced using standard flat--field normalisation and sky subtraction techniques. Wavelength calibration was achieved using ThAr comparison spectra and to an accuracy of less than 0.5 km/s. The radial velocities were calculated by fitting a Gaussian to the core of each line and values of $V_{r}$ can be found in Table 1.

To obtain the orbital solution we initially considered all of the radial velocity measurements. It was necessary to subtract centre of mass velocities ($\gamma$-velocities) for each source to account for differences between the data sets due to the various telescopes, instruments, choice of lines and methods of radial velocity determination used. The radial velocities of a  total of 48 different lines were used in the calculation of our ephemeris -- the majority of these can be found in Table 2 of Gies \& Bolton (1982); we also use additional He\,{\sc i} lines ($\lambda4921$, $\lambda5015$, $\lambda5047$, $\lambda5875$, $\lambda6678$), oxygen lines (O\,{\sc ii}$\lambda4349$, O\,{\sc ii}$\lambda4366$, O\,{\sc iii}$\lambda4650$), Mg\,{\sc ii}\,$\lambda4481$ and N\,{\sc ii},$\lambda4630$. As with previous authors, all lines were treated equally, although we note that inhomogeneities in the atmosphere of the supergiant (due to the variety of velocities in the stellar wind) cause greater shifting in the red HeI lines than the blue -- unfortunately it is not possible to correct for this without knowing exactly which authors used which lines for each radial velocity measurement. 

With so many radial velocities from fourteen different sources we were then able to omit those spectra for which O-C $> 20$ km/s and still have a total of 421 points from which to construct a radial velocity curve. We have also omitted the results of Seyfert \& Popper 1941 for reasons discussed in LaSala et al. (1998). The resultant radial velocity curve is shown in Figure 1; residuals are also plotted and show a definite lack of structure. Weighting all points by the inverse of their r.m.s. error and assuming an orbit of zero eccentricity the orbital parameters were obtained using the {\sc FOTEL}3 software. They can be found in Table 3, along with the elements calculated using the photometric data. We calculate the orbital period to be $P_{orb,sp}=5.599829\pm0.000016$ and define T$_0$ as the time of superior conjunction of the black hole.

Our new, high quality spectra have enabled us to extend the baseline for 
the ephemeris by over 400 days. As well as improving the accuracy of the 
radial velocity curve it also emphasizes the stability of the orbit of 
Cygnus X-1 over long time intervals.

\begin{table}
\caption{Radial velocities for Crimean observations of 1997}
\vspace*{0.5cm}
\begin{center}
\begin{tabular}{lcc}
\hline 
JDh (2400000+)&$V_{r}$ (km/s)&r.m.s. error\\
\hline
50615.4956 & -48.5  &1.0\\
50623.4675 &  66.4  &1.4 \\ 
50624.4585 & -14.3  &1.2 \\
50625.4553 & -63.3  &2.3  \\ 
50626.4792 & -72.7  &1.3 \\  
50648.4623 & -78.5  &1.3 \\ 
50649.5196 &  -7.5  &1.8 \\  
50650.5054 &  56.8  &1.3 \\     
50651.5033 &  62.9  &1.4  \\  
50653.5095 & -65.4  &1.1 \\  
50658.4643 & -26.1 & 1.0\\
50660.3914 & -52.1 &1.6\\
50661.3882 &  33.5 &1.7  \\
50663.4452 &   4.2 &1.0\\ 
50667.4747 &  77.3 &1.6 \\   
50668.4164 &  66.2 &1.7 \\
50669.4382 &  10.0 &2.5\\
50675.3550 & -44.7 &1.2\\   
50676.4727 & -86.0 &1.4 \\    
50677.4751 & -28.8 &1.4 \\
\end{tabular}
\end{center}
\end{table}  
\vspace*{-0.5cm}

\begin{table*} 
 \caption[ ]{Sources of RV velocities for Cygnus X-1}
\begin{center}
\vspace*{0.5cm}
\begin{tabular}{lcccc} 
\hline  
JD (2400000+)&Source&No. of spectra&$\gamma$-velocity (km/s)&O-C (km/s)\\
\hline
41159-41588&Smith et al., 1973&9&0.1$\pm$2.2&6.3\\
41163-41255&Webster, Murdin, 1972&16&0.1$\pm2.7$&10.5\\
41213-44795&Gies, Bolton, 1982&78&-1.8$\pm$0.8&6.6\\
41214-41477&Brucato, Kristian, 1973 &12&1.1$\pm$1.7&5.8\\
41269-41012&Mason et al., 1974&14&2.5$\pm$2.3&8.7\\
41515-42670&Walker, Yang, Glaspey, 1978&13&-0.6$\pm$1.0&3.7\\
41844-41290&Brucato, Zappala, 1974&17&-3.9$\pm$1.8&7.6\\
42205-42910&Abt, Hitzen, Levi, 1977&79&-7.9$\pm$1.1&7.9\\
43090-44768&Aab, 1983b&24&-6.3$\pm$1.3&5.9\\
44513-45895&Ninkov et al., 1987&84&-5.6$\pm$0.7&6.6\\
46332-46635&Sowers et al., 1998&14&-9.8$\pm$1.4&5.1\\
49217-49538&Canalizo et al., 1995&6&10.6$\pm$1.9&3.6\\
50228-50255&LaSala et al., 1998&33&-0.9$\pm$1.8&10.2\\
50615-50677&Crimea 1997 (this work)&20&-4.1$\pm$1.9&8.4\\

\hline
\end{tabular}
\end{center}
\end{table*}

\begin{figure*}
  \resizebox{\hsize}{!}{\includegraphics{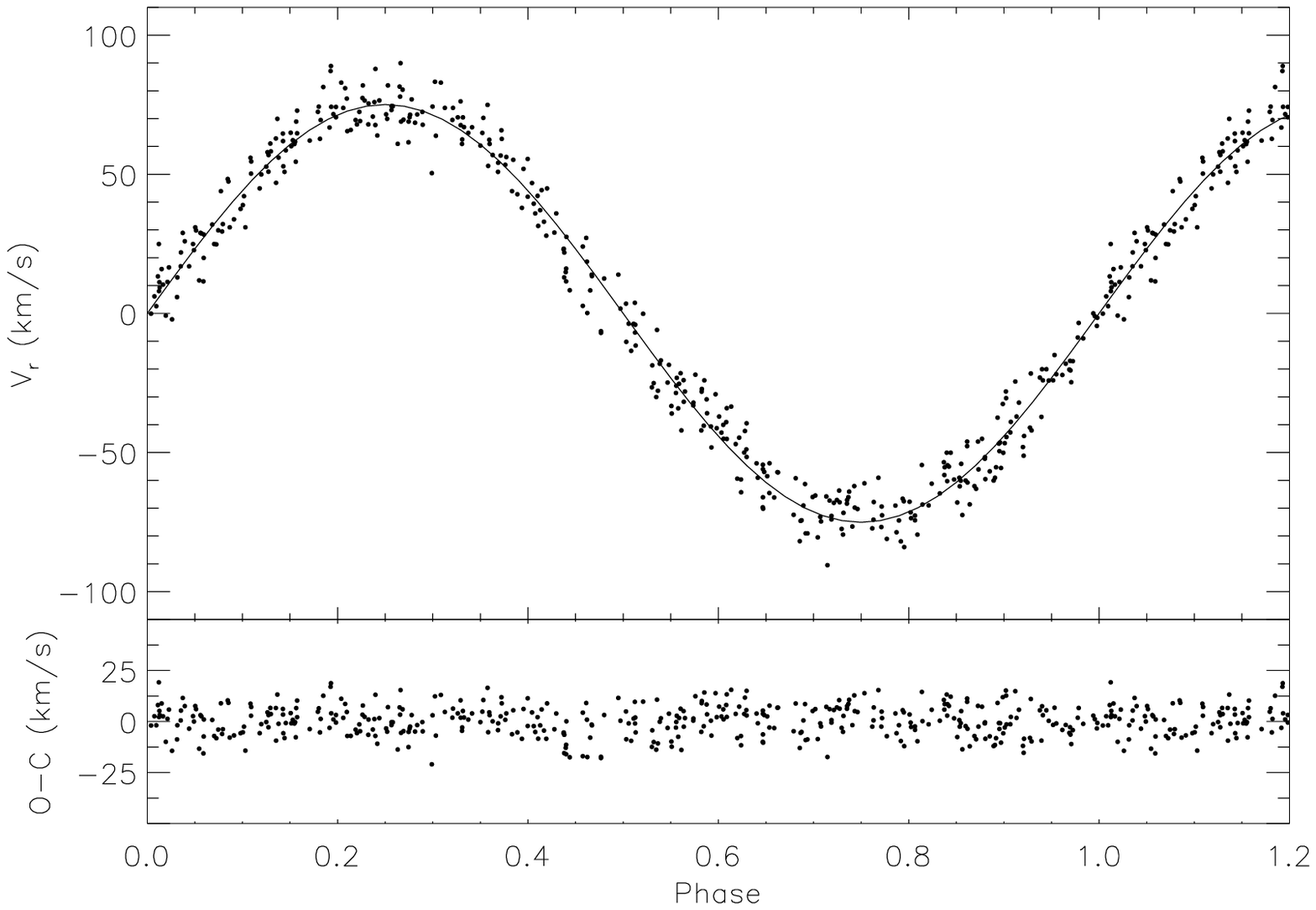}}
  \caption{Top: Radial velocity curve using accumulated data compared with theoretical curve from Balog, Goncharskij \& Cherepashchuk (1981), Bottom: Residuals}
  \label{fig:Radial Velocity Curve}
\end{figure*}

\subsection{Photometry}

The photometric {\it UBV} data were obtained on the 60cm telescope at the Crimean Laboratory (Nauchny, Crimea) of the Sternberg Astronomical Institute during 1971--1997. The pulse-counting photometer was used except for eight nights in July--August 1997 when it was changed to a CCD photometer with ST6V Camera (SBIG). All observations were made with reference to the standard BD $+35^{\circ}3816$, $V=9^m.976, B-V=0^m.590, U-B=0^m.064$ (Lyuty 1972). We therefore have a homogeneous photometric data set containing more than 800 observtions. For this reason we did not use the photometric data of other authors except those of Khaliullin \& Khaliullina (1981) which were obtained with the same telescope, photometer and local standard.

Using a DFT we obtained the photometric period in each of the {\it U, B} and {\it V} filters. The mean for the three filters, $P_{orb,phot}=5.599836\pm0.000024$ days, is in remarkable agreement with the spectroscopic period. We do not present the power spectrum here as the peak corresponding to the photometric period (and its aliases) is the only significant peak in the frequency range from zero up to the Nyquist frequency.

Previously Kemp et al. (1987) gave the most accurate value for the photometric period and the epoch of primary minimum (superior conjunction of X-ray source): \\$\mbox{Min I} = \mbox{JD}2441163.631(\pm 0.005) + 5.59985(\pm 0.00012)$. The low accuracy of the period value may be due to combining different data sets. However, using the same data Lloyd \& Walker (1989) obtained $P_{orb}=5.59982\pm0.00005$, which coincides with the more accurate value of $5.59982(\pm.00004)$ (Voloshina, Lyuty \& Tarasov 1997). Nadzhip et al. (1996) also improved the epoch of primary minimum to\\ JD 2441163.547$\pm0.005$.

Our new epoch of superior conjunction (Table 3) corresponds to primary minimum at $\mbox{JD}2441163.529\pm 0.009$, so, we can give the most accurate ephemeris as:

\hspace*{-0.5cm}$\mbox{Min I} = \mbox{JD}2441163.529(\pm 0.009) + 5.599829(\pm 0.000016)E$

\vspace*{0.075cm}
Using this ephemeris we have constructed the mean $UBV$ curves for our 27 years' worth of photometric data (Figure 2). To increase the homogeneity of the data set we have not used all available measurements, but only 1-5 per night, totalling 827 $UBV$ measurements. The observed mean curves are compared with the theoretical ellipsoidal curves of Balog, Goncharskij \& Cherepashchuk (1981), for which  $i=50^{\circ}, q=0.33, \mu=0.9, T_{\ast}=30000 K$.

The discrepancy between the observed mean light curves and the theoretical curve is probably due to the stellar wind. Were the accretion disk to be responsible then we would expect to see the maximum additional emission in the $V$-band (assuming disk temperature of $\sim10000$ K) at inferior conjunction of the black hole (phase 0.5). Clearly this is not the case, the maximum additional emission being in the $U$-band at phase 0.05 and 0.5. Further analysis of the photometry is in preparation.

\begin{figure}
  \resizebox{\hsize}{!}{\includegraphics{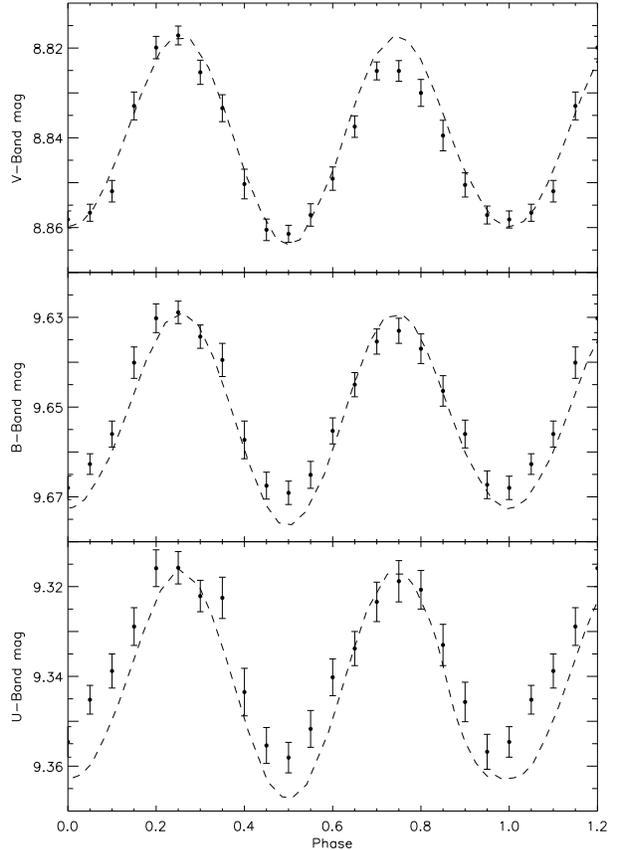}}
  \caption{Observed mean lightcurves for 1971-1997 compared with theoretical ellipsoidal curve}
  \label{fig:Mean Lightcurves}
\end{figure}

\vspace*{-0.5cm}
\section{Conclusion}

We have refined the orbital period of Cygnus X-1 and remain within agreement of the results of many other authors. Our value of 5.599829 days for the orbital period also provides an accurate fit to 27 years' worth of photometry. Further mutiwavelength analysis will follow in another paper.

\begin{table}
  \caption[ ]{The Orbital Elements of Cygnus X-1}
\vspace{0.25cm}
\begin{tabular}{lcc}
\hline
\hline
Element&Spectroscopic&Photometric\\
\hline
Period (days)&5.599829$\pm$0.000016&5.599836$\pm$0.000037\\
T$_0$ (JDh)&2441874.707$\pm$0.009&2441163.529$\pm$0.009\\
$K_1$ (km/s)&74.93$\pm$0.56&--\\
$f(M)  (M_{\odot}$)&0.244&--\\
\hline
\end{tabular}
\end{table}

\begin{acknowledgements}
We gratefully acknowledge receipt of financial support from the Royal Society for collaborative work with the Former Soviet Union. C. Brocksopp acknowledges a PPARC studentship; V.Lyuty and A.Tarasov are grateful to RFBR (grant 98-02-17067) for support in part. The majority of this work was accomplished on the Sussex {\sc STARLINK} node. 

\end{acknowledgements}

\end{document}